\documentclass[10pt]{article}
\usepackage[utf8]{inputenc}
\usepackage[T1]{fontenc}
\usepackage{graphicx}
\usepackage{amsmath,amsfonts,amssymb,bm}
\usepackage{siunitx}
\usepackage{booktabs}
\usepackage{setspace}
\usepackage{authblk}
\usepackage{caption}
\usepackage{float}
\usepackage{cite}
\usepackage{subcaption}
\usepackage{hyperref}
\usepackage[left=2.0cm,right=2.0cm,top=2cm,bottom=2cm]{geometry}

\hypersetup{colorlinks=true, linkcolor=blue, citecolor=blue, urlcolor=blue}

\newcommand{\bk}{\mathbf{k}}
\newcommand{\bs}{\hat{\mathbf{s}}}
\newcommand{\bE}{\mathbf{E}}
\newcommand{\bI}{\mathbf{I}}
\newcommand{\brho}{\boldsymbol{\rho}}
\newcommand{\bR}{\mathbf{R}}
\newcommand{\avg}[1]{\left\langle #1 \right\rangle}
\newcommand{\doi}[1]{\href{https://doi.org/#1}{doi:#1}}
\newcommand{\tr}{\operatorname{tr}}

\title{\textbf{Vectorial Control of Scintillation in Kolmogorov Turbulence and Quality Parametrization}}

\author[2,*]{Shouvik Sadhukhan}
\affil[2]{\small Applied and Adaptive Optics Laboratory, Department of Physics, Indian Institute of Space Science and Technology (IIST), Valiamala, Thiruvananthapuram 695547, Kerala, India. Email: shouvikphysics1996@gmail.com}
\affil[*]{\small Corresponding author: shouvikphysics1996@gmail.com}

\date{\today}

\begin{document}
\maketitle

% =====================================================================
\begin{abstract}
Intensity scintillation limits free-space optical links in strong turbulence. We examine experimentally how sequential linear polarization filtering modifies the scintillation of a laser beam distorted by Kolmogorov-type turbulence, and interpret the results with a vector-field and fourth-order-moment model. A collimated He--Ne beam (632.8~nm) crossed a rotating pseudo-random phase plate (Fried parameter $r_0\approx0.6$~mm) and zero to five co-aligned thin-film polarizers; 200 camera frames were recorded per configuration, together with a turbulence-free reference. For fully developed speckle the model gives a total-field index $\sigma_I^2=(1+P^2)/2$ in terms of the degree of polarization $P$. From zero to five polarizers, the Gaussian-envelope scintillation index fell from 0.083 to 0.002 and the single-pixel index from 0.80 to 0.27 (about 66\%), whereas beam wander remained comparable across configurations. As Gaussian statistics alone cannot explain this reduction, we attribute it to rejection of a depolarized, strongly fluctuating component and derive a quantitative criterion for this mechanism. Passive polarization filtering thus offers a simple, low-cost route to scintillation mitigation in free-space optical terminals.\\[4pt]
\end{abstract}

% =====================================================================
\section{Introduction}\label{sec:intro}

Free-space optical (FSO) communication offers licence-free, high-capacity links for terrestrial, airborne and satellite platforms, yet its reliability is governed by optical turbulence \cite{andrews,zhu,khalighi,kaushal}. Refractive-index fluctuations obeying Kolmogorov statistics \cite{kolmogorov,fried,noll} imprint random phase and amplitude on the wavefront, producing beam wander, speckle and irradiance fluctuations quantified by the scintillation index, whose probability distribution in strong turbulence departs markedly from log-normal \cite{alhabash}.

Established countermeasures include receiver aperture averaging, reduced transmitter coherence \cite{ricklin,gbur2002,dogariu,gu}, active wavefront correction and wavefront shaping \cite{roddier,vellekoop}, and structured or vortex beams, whose orbital-angular-momentum modes are nevertheless distorted by turbulence \cite{paterson,willner}. Most of these require sensors, feedback loops or spatial light modulators, so passive alternatives remain attractive \cite{sadhukhan}.

Polarization is a comparatively underexploited degree of freedom. The unified theory of coherence and polarization \cite{wolfbook,mandelwolf,wolf2003,korotkova2004} and the matrix formalisms built upon it \cite{gori1998,korotkovagbur,laatikainen} describe partially polarized, partially coherent beams through second-order correlation matrices, whereas intensity fluctuations are fourth-order quantities. For Gaussian field statistics \cite{goodman,reed} the two orders are linked: Korotkova showed that the scintillation index of an electromagnetic beam depends on its state of polarization and that, for identical intensity profile and coherence, natural light scintillates half as strongly as fully polarized light \cite{korotkova2006,korotkova2006stokes,korotkova2008}. This prediction concerns the total intensity and is not by itself a recipe for suppression, because a polarizer removes light and reshapes the detected statistics. Moreover, when turbulence tilts the local wavefront the field must remain transverse to the local wave vector \cite{bornwolf}, which introduces polarization-dependent corrections absent from scalar treatments.

Here we study experimentally, and with a supporting vector model, how sequential polarization filtering modifies scintillation after strong turbulence emulated by a rotating pseudo-random phase plate. Section~\ref{sec:theory} develops the model from the experimental configuration, Section~\ref{sec:experiment} describes the apparatus and data analysis, Section~\ref{sec:results} presents and discusses the results, and Section~\ref{sec:conclusion} concludes.

% =====================================================================
\section{Theoretical Framework}\label{sec:theory}

\subsection{Experimental Context and Model Chain}

The measurement chain (Fig.~\ref{fig:setup}, Sec.~\ref{sec:experiment}) consists of (i) a collimated, $\hat{\mathbf x}$-polarized Gaussian beam, (ii) a thin Kolmogorov-type phase screen (the rotating pseudo-random phase plate, PRPP), (iii) $N$ co-aligned linear polarizers, (iv) free-space propagation over an effective distance $z$, and (v) intensity detection by a camera, with $N=0,\dots,5$. The model follows this order. At the plate plane ($z=0$) the input field is
\begin{equation}
\bE_{\rm in}(\brho)=\hat{\mathbf x}\,E_0\exp\!\left(-\frac{|\brho|^{2}}{w_0^{2}}\right),
\label{eq:input}
\end{equation}
with $w_0\simeq2.5$~mm. The plate imprints a random phase $\phi_t(\brho)$ whose statistics are Kolmogorov-like, with phase structure function $D_\phi(\rho)=6.88\,(\rho/r_0)^{5/3}$ and Fried parameter $r_0\approx0.6$~mm, so that $2w_0/r_0\approx8$. With $\kappa=2\pi/\lambda$, the accumulated phase is
\begin{equation}
\Phi(\brho,z)=\kappa z-\phi_t(\brho).
\label{eq:totalphase}
\end{equation}
A scalar thin-screen description, $\bE\to\bE\,e^{i\Phi}$, keeps both the propagation direction and the polarization unchanged. Because the plate has finite thickness and local slope, we instead treat it as a refracting surface $z_s=\phi_t/\kappa$ and impose Maxwell transversality explicitly.

\subsection{Local Wave Vector and Transversality Constraint}

The local wave vector is $\bk=\nabla\Phi=\kappa\,(-\varepsilon_x,-\varepsilon_y,1)$ with dimensionless slopes $\varepsilon_{x,y}=\partial_{x,y}\phi_t/\kappa$ and $\varepsilon^{2}=\varepsilon_x^{2}+\varepsilon_y^{2}$. The unit ray direction is $\bs=\bk/|\bk|=(-\varepsilon_x,-\varepsilon_y,1)/\sqrt{1+\varepsilon^{2}}$. Requiring $\bk\cdot\bE_t=0$, the field behind the plate is the transverse projection of the trial field,
\begin{equation}
\bE_t(\brho)=\left(\bI-\bs\bs^{\mathsf T}\right)\hat{\mathbf x}\,\mathcal E(\brho)
=\mathcal E(\brho)\,\mathbf A(\brho),\qquad
\mathcal E(\brho)=E_0\,e^{-|\brho|^{2}/w_0^{2}}\,e^{i\Phi(\brho,0)},
\label{eq:Et}
\end{equation}
with components
\begin{equation}
A_x=1-\frac{\varepsilon_x^{2}}{1+\varepsilon^{2}},\qquad
A_y=-\frac{\varepsilon_x\varepsilon_y}{1+\varepsilon^{2}},\qquad
A_z=\frac{\varepsilon_x}{1+\varepsilon^{2}}.
\label{eq:Acomp}
\end{equation}
The projection satisfies $|\mathbf A|^{2}=A_x\le1$ (Appendix~\ref{app:projection}), i.e.\ the transversality constraint removes the fraction $\varepsilon_x^{2}/(1+\varepsilon^{2})$ of the trial intensity. Thus $E_x$ remains dominant, $E_z\simeq\varepsilon_x E_x$ is the longitudinal field generated by local tilt, and $E_y\simeq-\varepsilon_x\varepsilon_y E_x$ is the cross-polarized field generated by the mixed slope.

\begin{figure}[t!]
\centering
\includegraphics[width=0.95\textwidth]{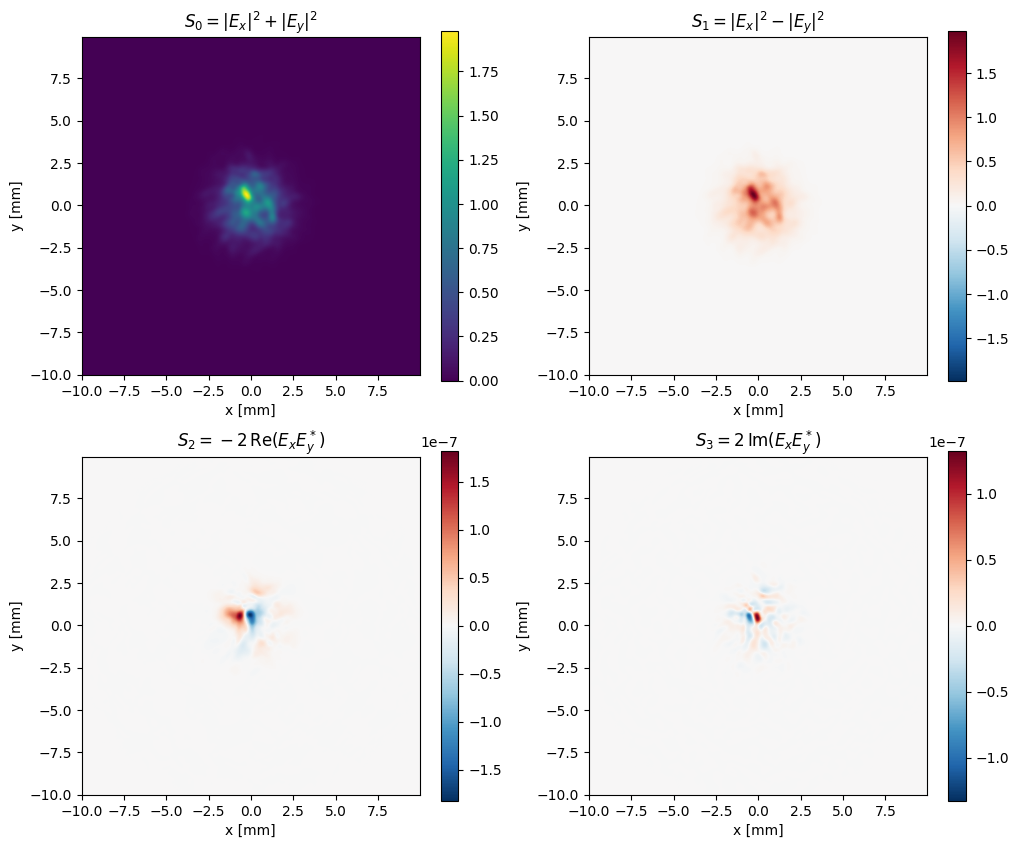}
\caption{Model Stokes-parameter maps at the detection plane for the turbulence-modified vector field of Eq.~\eqref{eq:Et}, before the analyzers. $S_0=|E_x|^2+|E_y|^2$, $S_1=|E_x|^2-|E_y|^2$, $S_2=-2\,\mathrm{Re}(E_xE_y^{*})$, $S_3=2\,\mathrm{Im}(E_xE_y^{*})$. Note the $10^{-7}$ scale of $S_2$ and $S_3$.}
% CONFIRM: numerical parameters (r0, w0, z, grid) used to generate this figure.
\label{fig:stokes}
\end{figure}

\emph{Order of magnitude.} From $D_\phi(r_0)^{1/2}/r_0\approx2.6/r_0$, the typical slope for $r_0=0.6$~mm is $\varepsilon\sim2.6/(\kappa r_0)\approx4\times10^{-4}$. The longitudinal intensity fraction is then $\sim\varepsilon^{2}\approx2\times10^{-7}$ and the cross-polarized intensity fraction is $\sim\varepsilon^{4}\approx10^{-13}$. Figure~\ref{fig:stokes} confirms that the model field stays essentially linearly polarized ($S_1\simeq S_0$, $|S_{2,3}|\sim10^{-7}S_0$). The geometric transversality coupling alone is therefore far below the dynamic range of an 8-bit camera ($\sim4\times10^{-3}$); any experimentally significant cross-polarized light must originate in the plate material, scattering, or other depolarizing elements. This point is quantified in Sec.~\ref{sec:twocomp}.

\subsection{Analyzers, Propagation and Detection}

Each thin-film polarizer is described by the Jones matrix $\mathbf J_p=\mathrm{diag}(1,\eta)$, where $\eta\ll1$ is the residual cross-polarized amplitude transmission, so that the extinction ratio is $\mathrm{ER}=\eta^{-2}$. For $N$ co-aligned polarizers in series, $\mathbf J_p^{N}=\mathrm{diag}(1,\eta^{N})$, and the cross-polarized leakage in intensity scales as $\mathrm{ER}^{-N}$ (multiple reflections between elements are neglected). Each transverse component propagates from the plate to the detection plane through the Fresnel integral
\begin{equation}
E_j(\bR,z)=\frac{e^{i\kappa z}}{i\lambda z}\iint \mathcal E(\brho)\,A_j(\brho)\,
\exp\!\left[\frac{i\kappa}{2z}\,|\bR-\brho|^{2}\right]d^{2}\rho,\qquad j=x,y,z,
\label{eq:fresnel}
\end{equation}
where an imaging lens maps the plane at distance $z$ onto the sensor. All components share one kernel; their difference resides in the source weights $A_j$, so turbulence-induced vector structure is set at the plate and merely transported by diffraction. The component-wise integrals are given in Appendix~\ref{app:fresnel}. The longitudinal field does not couple to a camera at near-normal incidence, so the recorded intensity after $N$ polarizers is
\begin{equation}
I_N(\bR)=|E_x(\bR,z)|^{2}+\eta^{2N}|E_y(\bR,z)|^{2}.
\label{eq:IN}
\end{equation}

\subsection{Fourth-Order Statistics and the Purity Relation}

Rotation of the plate presents successive realizations of $\phi_t$ to the beam. Replacing ensemble averages by temporal averages $\avg{\cdot}$ (ergodic assumption), the scintillation index at a detector point is $\sigma_I^{2}=\avg{I^{2}}/\avg{I}^{2}-1$. For the total field $I=I_x+I_y$, Korotkova's decomposition \cite{korotkova2008} reads
\begin{equation}
\sigma_I^{2}=\frac{\sigma_{xx}^{2}I_x^{2}+2\sigma_{xy}^{2}I_xI_y+\sigma_{yy}^{2}I_y^{2}}{(I_x+I_y)^{2}},\qquad
\sigma_{ij}^{2}=\frac{\Gamma^{(2,2)}_{iijj}-\Gamma^{(1,1)}_{ii}\Gamma^{(1,1)}_{jj}}{\Gamma^{(1,1)}_{ii}\Gamma^{(1,1)}_{jj}},
\label{eq:SI}
\end{equation}
with $\Gamma^{(1,1)}_{ij}=\avg{E_i^{*}E_j}$ and $\Gamma^{(2,2)}_{iijj}=\avg{I_iI_j}$. If the field components at a point are circular complex Gaussian (fully developed speckle) \cite{goodman,reed}, the fourth-order moments factorize into second-order ones and Eq.~\eqref{eq:SI} reduces to the \emph{purity relation} (Appendix~\ref{app:purity})
\begin{equation}
\sigma_I^{2}=\frac{\tr(\mathsf J^{2})}{[\tr\mathsf J]^{2}}=\frac{1+P^{2}}{2},\qquad
P=\sqrt{1-\frac{4\det\mathsf J}{[\tr\mathsf J]^{2}}},
\label{eq:purity}
\end{equation}
where $\mathsf J_{ij}=\avg{E_i^{*}E_j}$ is the polarization matrix and $P$ the degree of polarization. A natural beam ($P=0$) has $\sigma_I^{2}=1/2$ and a fully polarized beam ($P=1$) has $\sigma_I^{2}=1$. Two consequences guide the experiment. First, raising $P$ cannot lower the total-field index within Gaussian statistics. Second, after an ideal analyzer the camera records a single component with $\sigma_{xx}^{2}=1$. A reduction of the measured index by polarization filtering therefore requires a departure from fully developed Gaussian speckle.

\subsection{Partially Developed Speckle with a Depolarized Component}\label{sec:twocomp}

We consider a minimal two-component field, $E_x=s+g_x$ and $E_y=g_y$, where $s$ is a coherent co-polarized amplitude of power $S=|s|^{2}$ and $g_{x,y}$ are independent circular Gaussian fields (diffuse light) of powers $d_x,d_y$. Then (Appendix~\ref{app:twocomp})
\begin{equation}
\sigma_{\rm tot}^{2}=\frac{d_x^{2}+d_y^{2}+2Sd_x}{(S+d_x+d_y)^{2}},\qquad
\sigma_{x}^{2}=\frac{d_x(d_x+2S)}{(S+d_x)^{2}},
\label{eq:twocomp}
\end{equation}
for the unfiltered and ideally analyzed intensities, and $\sigma_x^{2}<\sigma_{\rm tot}^{2}$ if and only if
\begin{equation}
d_y>\frac{2\,d_x\,(S+d_x)(2S+d_x)}{S^{2}}.
\label{eq:criterion}
\end{equation}
Filtering therefore lowers the scintillation index only when the strongly fluctuating diffuse light is predominantly cross-polarized relative to the co-polarized diffuse light, while a coherent co-polarized core survives. With $N$ imperfect polarizers $d_y\to\eta^{2N}d_y$ in Eq.~\eqref{eq:twocomp}, so the index approaches $\sigma_x^{2}$ monotonically as $N$ increases and saturates once $\eta^{2N}d_y\ll d_x$. This is the mechanism tested in Sec.~\ref{sec:results}; it is a testable hypothesis, not a result of the present measurements.

% =====================================================================
\section{Experimental Setup and Data Analysis}\label{sec:experiment}

\subsection{Optical Configuration}

\begin{figure}[t!]
\centering
\includegraphics[width=0.95\textwidth]{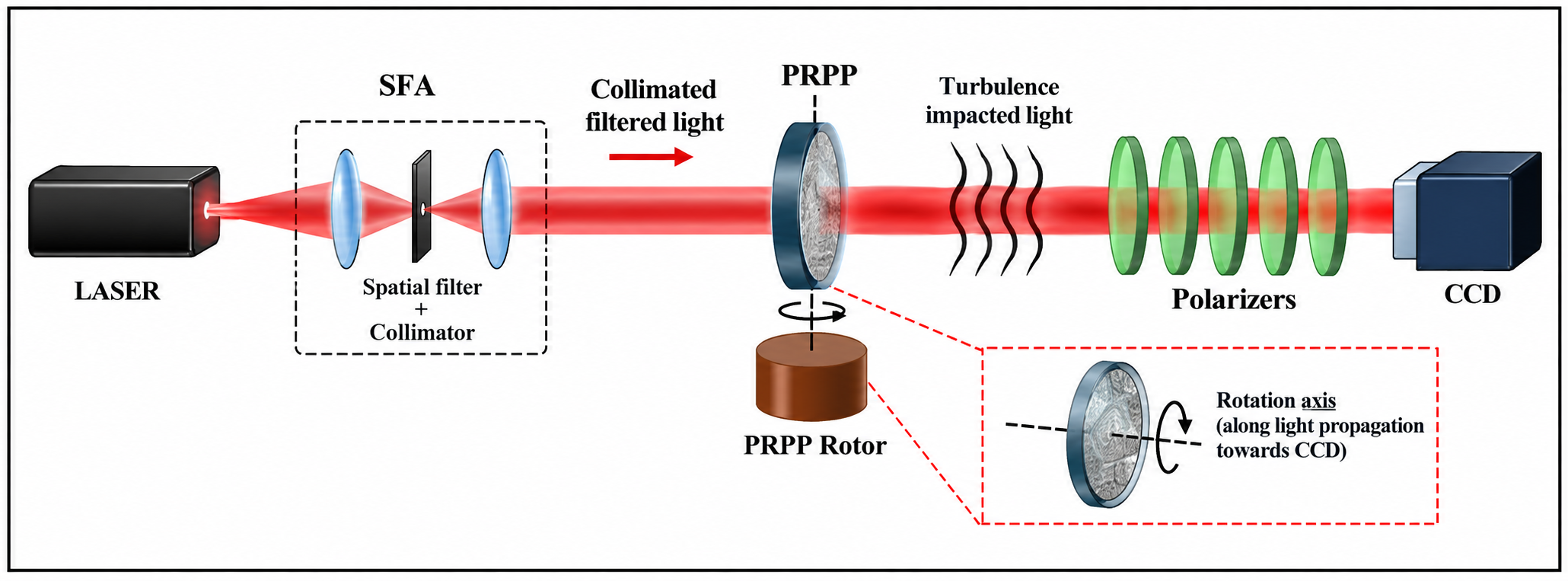}
\caption{Experimental set-up. A spatially filtered and collimated He--Ne beam crosses the rotating PRPP, 0--5 thin-film polarizers, and is recorded by the camera. Rotation of the PRPP provides new turbulence realizations over the 200 frames of each set.}
\label{fig:setup}
\end{figure}

The set-up is shown in Fig.~\ref{fig:setup}. A continuous-wave He--Ne laser ($\lambda=632.8$~nm) was spatially filtered and collimated by a spatial filter assembly (SFA: microscope objective, pinhole and collimating lens), yielding a near-diffraction-limited Gaussian beam of about 5~mm diameter ($1/e^{2}$). The beam traversed a pseudo-random phase plate (PRPP, Thorlabs EDU-RPP1) on a motorized rotation stage and then $N=0$--5 identical, co-aligned thin-film linear polarizers in series. An imaging lens relayed the output to a monochrome 8-bit camera (Thorlabs DCC1545M, $1280\times1024$ pixels, 5.2~\textmu m pitch; effective sampling 4.28~\textmu m after magnification, about 600 pixels across the FWHM). No adaptive-optics or wavefront-correction element was used, so the configuration is purely passive. The turbulence-free reference (Set~7) was recorded with the PRPP removed and all other components in place.

\subsection{Turbulence Emulation}

The PRPP comprises two BK7 windows enclosing an acrylic layer carrying a Kolmogorov-type phase profile, index-matched by polymer layers. Its nominal Fried parameter is $r_0\approx0.6$~mm, giving $2w_0/r_0\approx8$ across the beam. The plate rotated at 1~rpm. The equivalence with atmospheric propagation follows from the uniform-path plane-wave Fried parameter $r_0=[0.423\,\kappa^{2}C_n^{2}L]^{-3/5}$. For $L=1.5$~cm this requires $C_n^{2}\approx3.7\times10^{-7}~\mathrm{m^{-2/3}}$, many orders of magnitude above atmospheric values; conversely, for a representative strong near-ground value $C_n^{2}=10^{-14}~\mathrm{m^{-2/3}}$ the same $r_0$ is reached after $L\approx5.6\times10^{2}$~km. Matching $r_0$ reproduces the transverse coherence scale but not every propagation parameter (for example the Rytov variance), so the plate emulates strong-turbulence wavefront statistics rather than a specific link.

\subsection{Acquisition Protocol}

Seven sets were recorded (Table~\ref{tab:sets}), each of 200 frames at 5~fps with 10~ms exposure. At 1~rpm the 40~s record spans about $240^{\circ}$ of plate rotation, so no turbulence realization repeats within a set, while adjacent frames ($1.2^{\circ}$ apart) remain correlated (Sec.~\ref{sec:limits}).
% CONFIRM: identical exposure/gain and camera settings for all sets (see beam-volume discussion in Sec. 4.2).

\begin{table}[t!]
\centering
\caption{Experimental sets (200 frames each).}
\label{tab:sets}
\begin{tabular}{@{}lcl@{}}
\toprule
Set & Polarizers $N$ & Configuration \\
\midrule
1 & 0 & PRPP only (unmitigated turbulence)\\
2--6 & 1--5 & PRPP followed by $N$ co-aligned thin-film polarizers\\
7 & 0 & Reference: no PRPP, no polarizer\\
\bottomrule
\end{tabular}
\end{table}

\subsection{Data Analysis}

\emph{Single-pixel statistics.} The pixel of maximum intensity in frame~0 of each set was tracked through all 200 frames, giving a record $I_k$ in normalized units. The windowed scintillation index
\begin{equation}
\sigma_{I,k}^{2}=\frac{\avg{I^{2}}_{M}-\avg{I}_{M}^{2}}{\avg{I}_{M}^{2}},\qquad M=10~\text{frames},
\label{eq:window}
\end{equation}
was evaluated over a sliding window, and the mean and standard deviation over $k$ are reported. A record-level index $(\sigma/\mu)^{2}$ was also obtained from the mean and standard deviation of the full 200-frame record. Percentage reductions are quoted relative to Set~1, with the standard error of the mean (SEM) as uncertainty.

\emph{Gaussian-envelope statistics.} Each frame was fitted by nonlinear least squares with a two-dimensional elliptical Gaussian including Gram--Charlier/Edgeworth corrections, yielding centroid $(\mu_x,\mu_y)$, widths $(\sigma_x,\sigma_y)$, cross-correlation $\sigma_{xy}$ and integrated volume. The envelope scintillation index $\bar\sigma^{2}_{I}=N_f^{-1}\sum_k\sigma^{2}_{I,k}$ is the frame average of the index of the fitted peak amplitude, evaluated with the window of Eq.~\eqref{eq:window}.
% CONFIRM: definition of the Gaussian-fitted per-frame index sigma^2_{I,k} (here: fitted peak amplitude, same 10-frame window).

% =====================================================================
\section{Results and Discussion}\label{sec:results}

\subsection{Beam Morphology and Wander}

\begin{figure}[t!]
\centering
\includegraphics[width=0.95\textwidth]{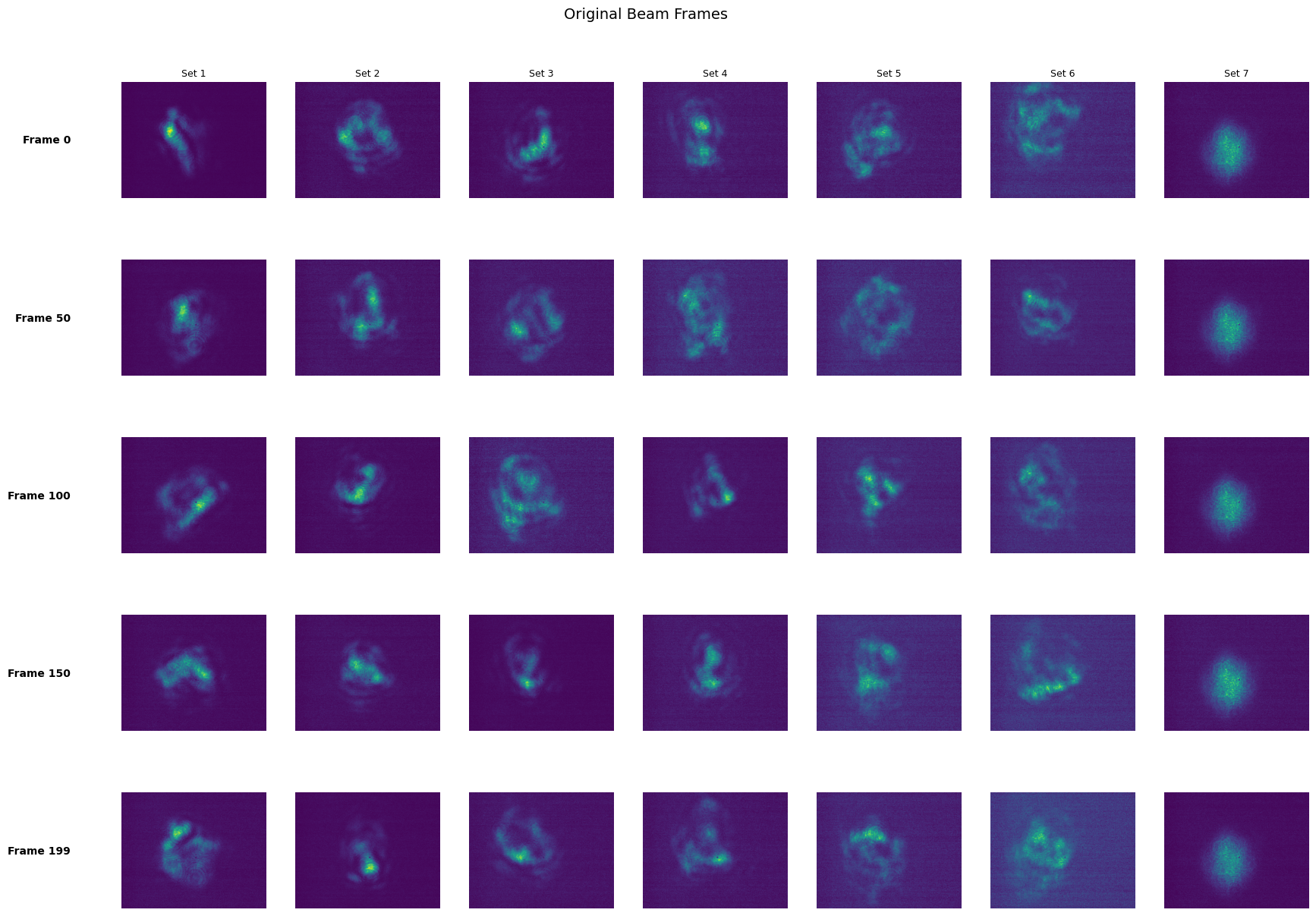}
\caption{Raw camera frames (frames 0, 50, 100, 150, 199) for Sets~1--7. Set~1 shows strongly distorted, randomly varying patterns, Sets~2--6 progressively broader and more filled patterns, and Set~7 the turbulence-free reference.}
\label{fig:raw}
\end{figure}

\begin{figure}[t!]
\centering
\includegraphics[width=0.95\textwidth]{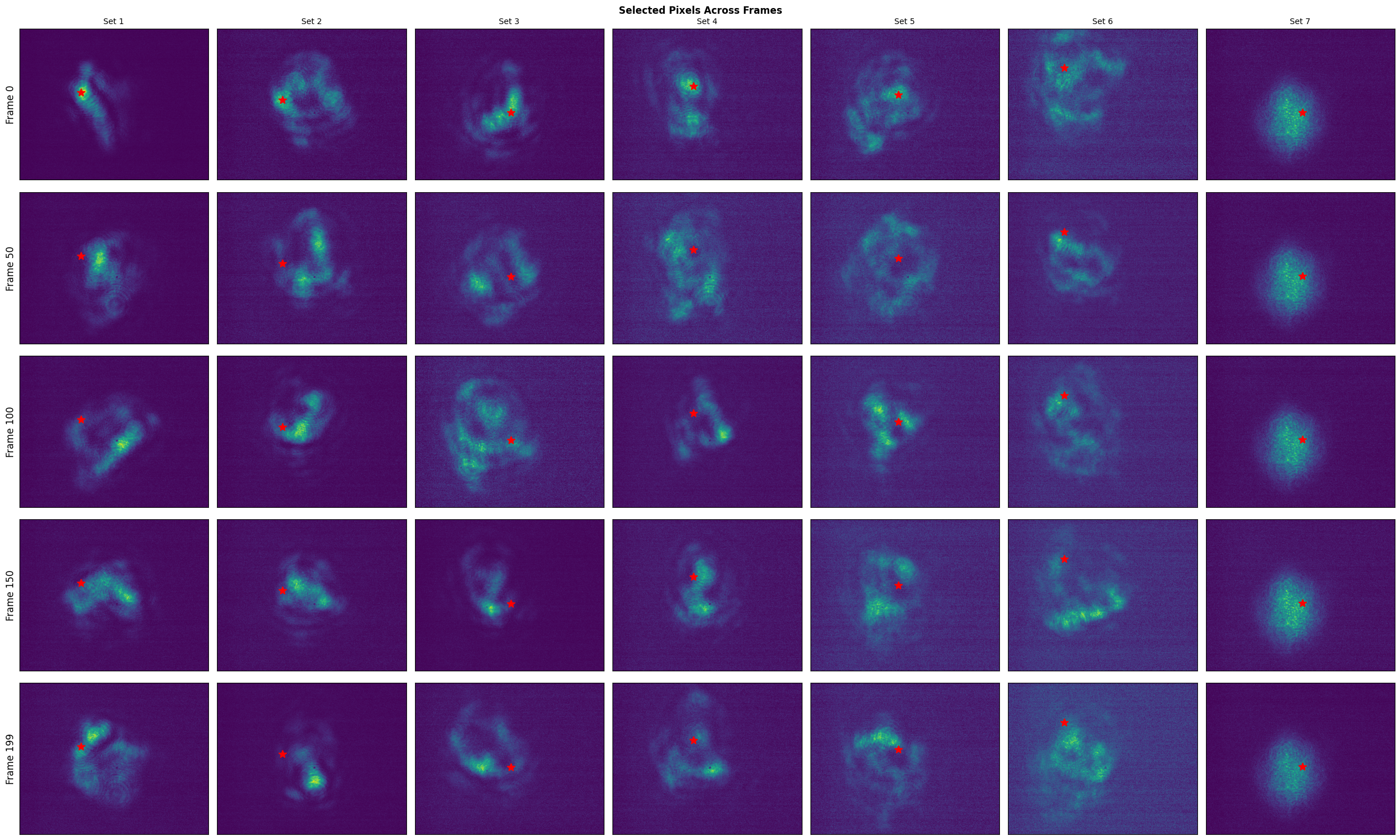}
\caption{Tracking of the pixel that is brightest in frame~0 (red marker) across the 200 frames of Sets~1--7.}
\label{fig:pointed}
\end{figure}

Without polarizers (Set~1) the beam is fragmented into irregular speckle that wanders across the sensor (Fig.~\ref{fig:raw}), whereas the reference (Set~7) is a stable, smooth Gaussian. With increasing $N$ the patterns become broader and more uniformly filled and their frame-to-frame variation decreases, although speckle structure persists and none of Sets~2--6 attains the reference profile. Tracking the frame-0 peak pixel (Fig.~\ref{fig:pointed}) shows displacements of hundreds of pixels in Set~1, progressively less migration of the bright region in Sets~2--6, and an essentially stationary pixel in Set~7.

\subsection{Fitted Beam Parameters}

\begin{figure}[t!]
\centering
\includegraphics[width=0.95\textwidth]{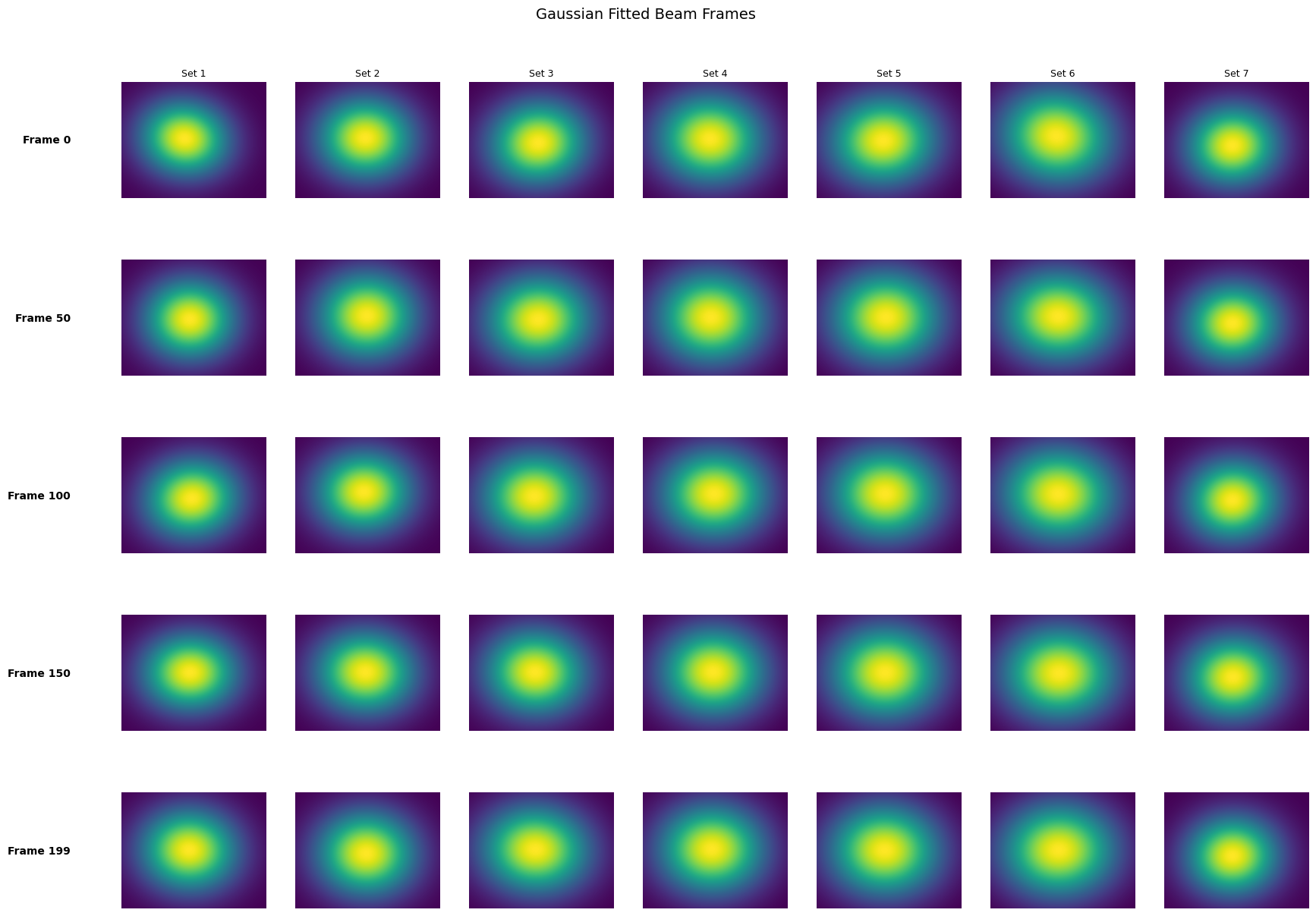}
\caption{Elliptical Gaussian fits to the frames of Fig.~\ref{fig:raw}.}
\label{fig:fitted}
\end{figure}

\begin{figure}[t!]
\centering
\includegraphics[width=0.95\textwidth]{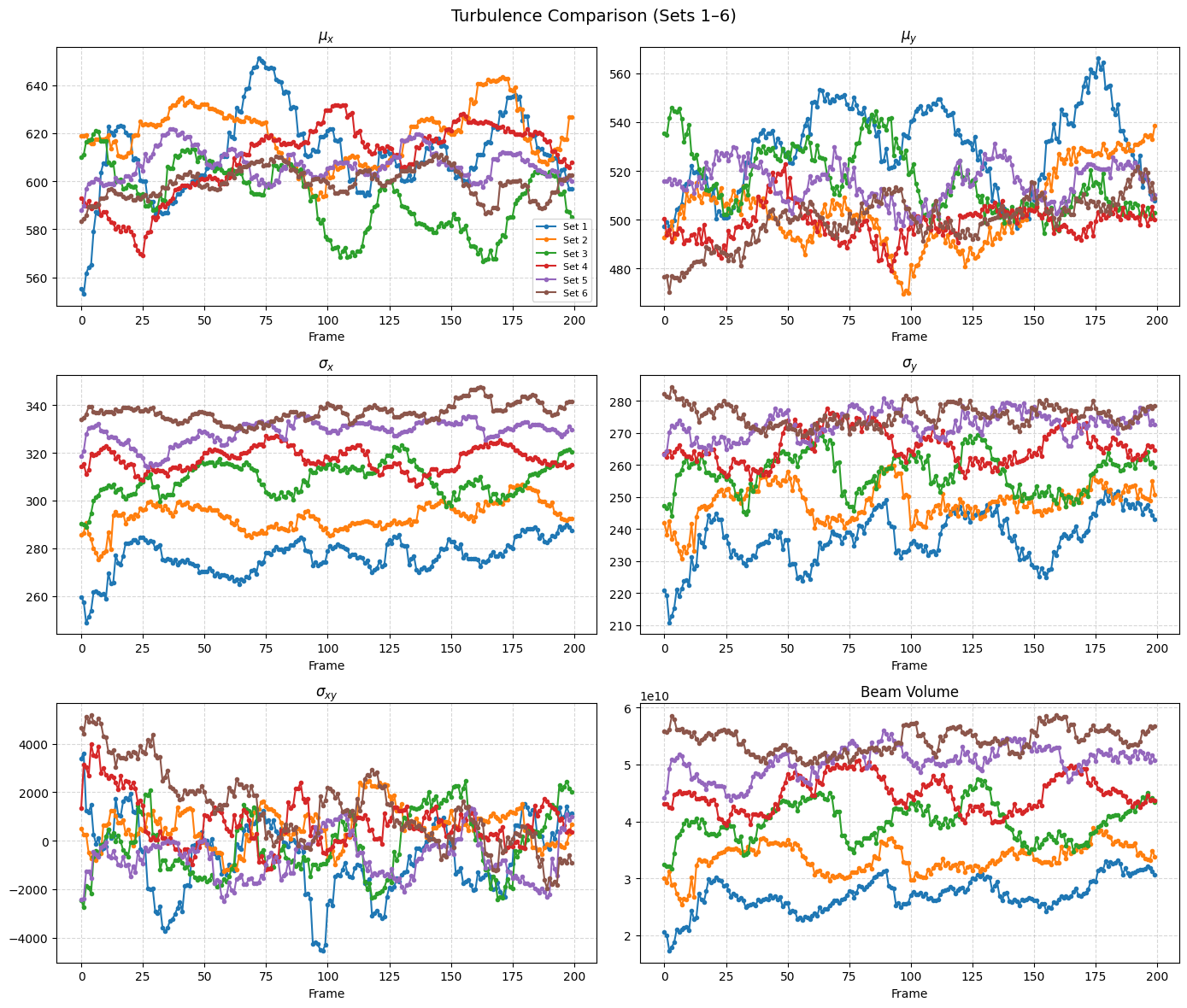}
\caption{Time series of fitted parameters for Sets~1--6: centroids $\mu_x,\mu_y$ (top), widths $\sigma_x,\sigma_y$ (middle), cross-correlation $\sigma_{xy}$ and volume (bottom).}
\label{fig:gauss_stats}
\end{figure}

The Gaussian fits (Fig.~\ref{fig:fitted}) provide a smooth envelope that is insensitive to speckle and sensor noise. Their time series (Fig.~\ref{fig:gauss_stats}) show the following. (i)~Centroid excursions of several tens of pixels are present in all sets with comparable amplitude, so polarizers do not reduce wander on the sampled time scales, as expected for a second-order (tilt) effect. (ii)~The fitted widths grow with $N$, from $\sigma_x\approx275$~px and $\sigma_y\approx235$~px (Set~1) to about 340~px and 275~px (Set~6): the speckle-fragmented beam is described by narrow sub-structures, which merge into a broader envelope as the polarizers are added. (iii)~$\sigma_{xy}$ fluctuates about zero with decreasing amplitude, so the ellipse axes are not systematically tilted. (iv)~The fitted volume, in raw camera counts, increases from about $2.5\times10^{10}$ (Set~1) to about $5.5\times10^{10}$ (Set~6). Since passive polarizers cannot add power, this trend indicates that acquisition conditions were not identical across sets; absolute volumes are therefore not used for inter-set comparison.

\subsection{Envelope Scintillation Index}

\begin{figure}[t!]
\centering
\includegraphics[width=0.85\textwidth]{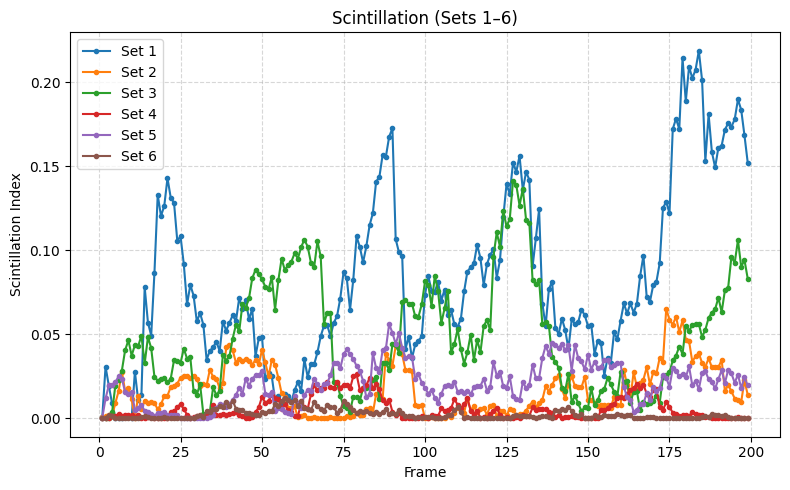}
\caption{Frame-resolved Gaussian-envelope scintillation index for Sets~1--6.}
\label{fig:scatter_power}
\end{figure}

\begin{figure}[t!]
\centering
\includegraphics[width=0.75\textwidth]{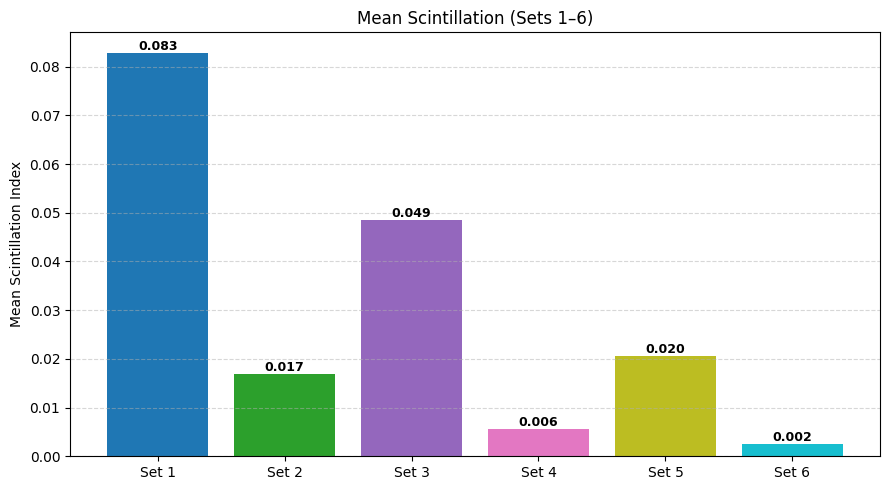}
\caption{Mean Gaussian-envelope scintillation index for Sets~1--6 (0.083, 0.017, 0.049, 0.006, 0.020, 0.002).}
\label{fig:bar_power}
\end{figure}

Figure~\ref{fig:scatter_power} shows the frame-resolved envelope index. Set~1 displays repeated bursts up to about 0.2, whereas Sets~4 and 6 stay close to zero. The set averages (Fig.~\ref{fig:bar_power}) are $0.083$, $0.017$, $0.049$, $0.006$, $0.020$ and $0.002$ for $N=0$--5, a reduction by a factor of about 40 between Sets~1 and 6 (the Set~6 value is quoted to one significant figure, so the factor is uncertain at the tens-of-percent level). The decrease is not monotonic: Sets~3 and 5 exceed their predecessors. Because this metric responds mainly to envelope shape and centroid dynamics, we regard it as a secondary indicator.

\subsection{Single-Pixel Records}

\begin{figure}[t!]
\centering
\includegraphics[width=0.62\textwidth]{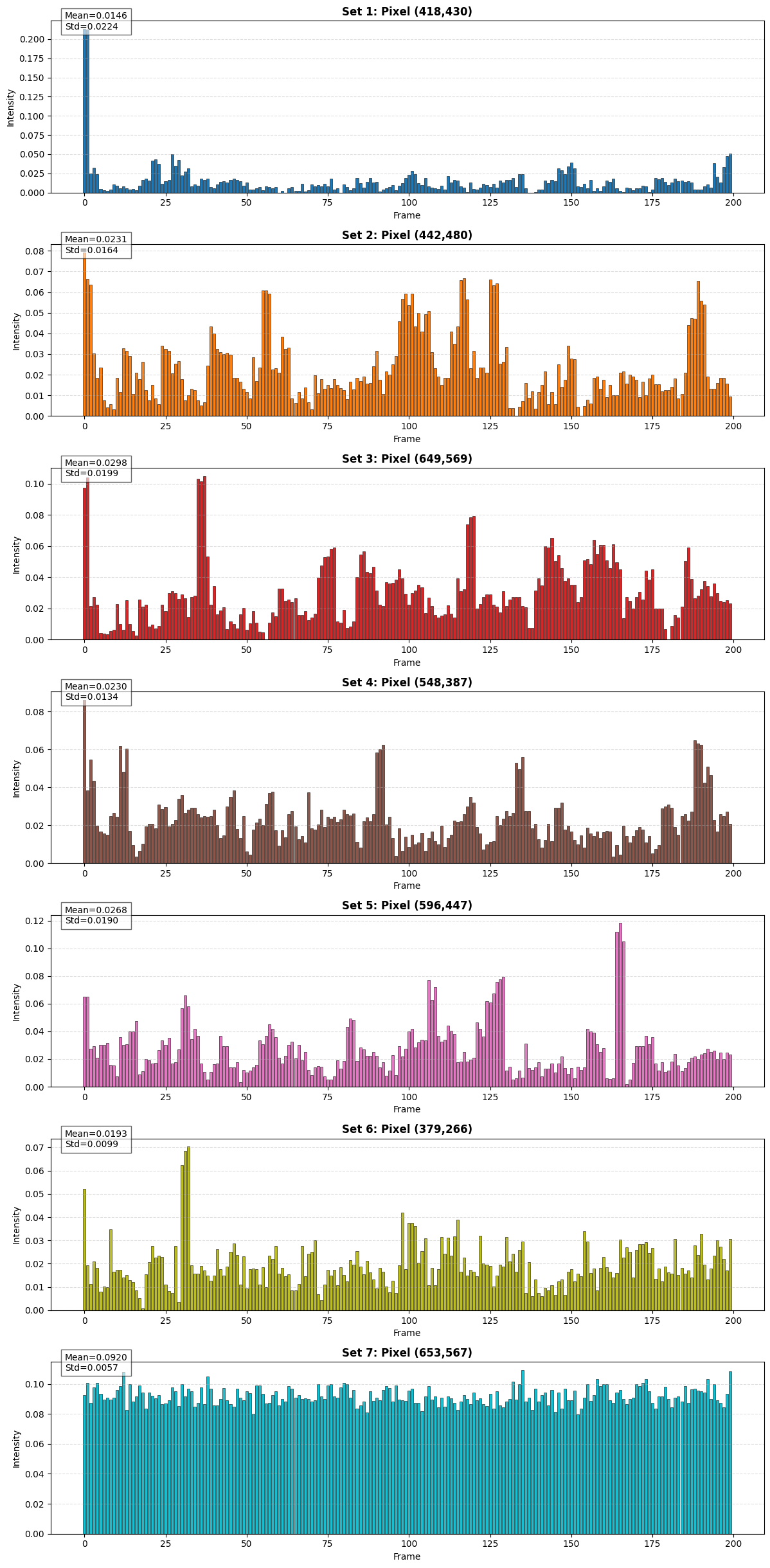}
\caption{Normalized intensity of the tracked pixel versus frame for Sets~1--7, with mean and standard deviation annotated.}
\label{fig:histogram}
\end{figure}

Figure~\ref{fig:histogram} gives the intensity of the tracked pixel versus frame. The Set~1 record is intermittent: after a large value at frame~0 (where the pixel was selected) it is mostly low with bursts as speckle and wander cross the pixel. Set~7 is flat at $\mu=0.0920$ with $\sigma=0.0057$. The records of Sets~2--6 are progressively more compact; Set~6 lies at $\mu=0.0193$, $\sigma=0.0099$ with one isolated excursion near frame~30. The record-level indices derived from the annotated moments (Table~\ref{tab:summary}) are 2.35 (Set~1), 0.50, 0.45, 0.34, 0.50 and 0.26 (Set~6), and $3.8\times10^{-3}$ for the reference. The standard deviation therefore decreases overall, but not monotonically (Set~5 exceeds Set~4). A record-level index above unity for Set~1 reflects non-stationary, intermittent illumination of a fixed pixel rather than saturated Gaussian speckle.

\subsection{Pixel-Level Scintillation Index}

\begin{figure}[t!]
\centering
\includegraphics[width=0.9\textwidth]{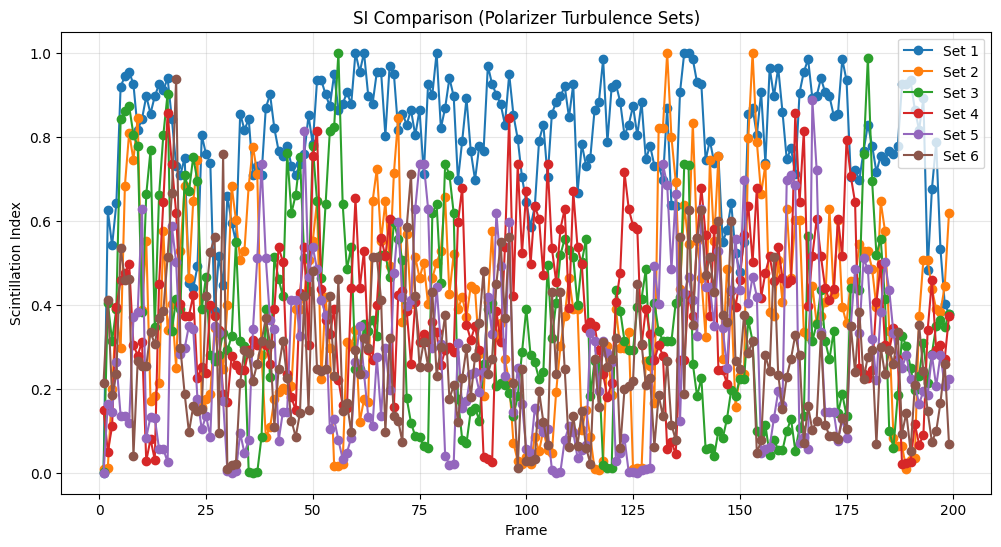}
\caption{Windowed pixel-level scintillation index (10-frame window) versus frame for Sets~1--6.}
\label{fig:SI_scatter}
\end{figure}

\begin{figure}[t!]
\centering
\includegraphics[width=0.75\textwidth]{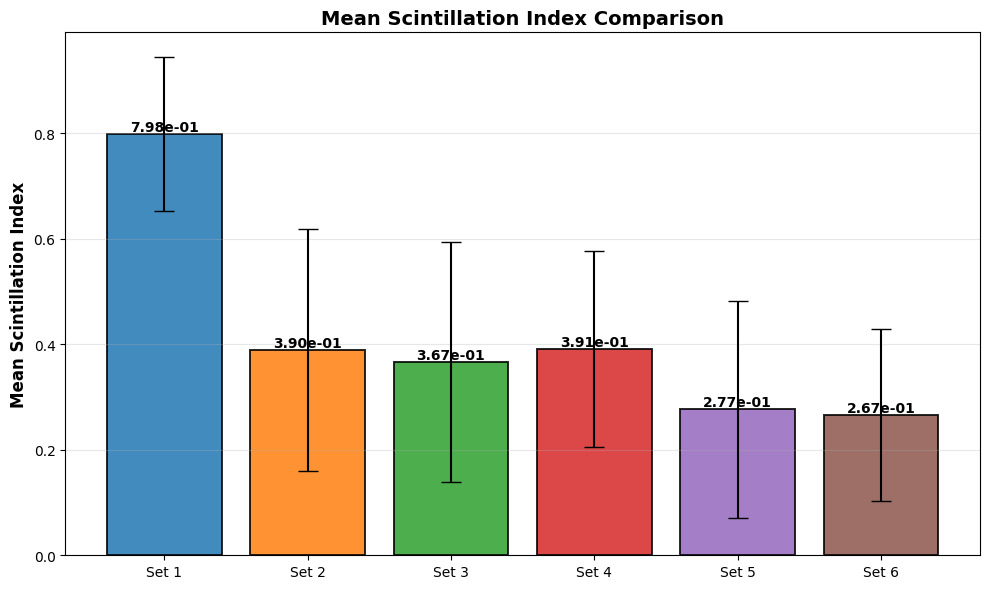}
\caption{Mean pixel-level scintillation index with standard-deviation bars for Sets~1--6.}
\label{fig:SI_bar}
\end{figure}

\begin{figure}[t!]
\centering
\includegraphics[width=0.95\textwidth]{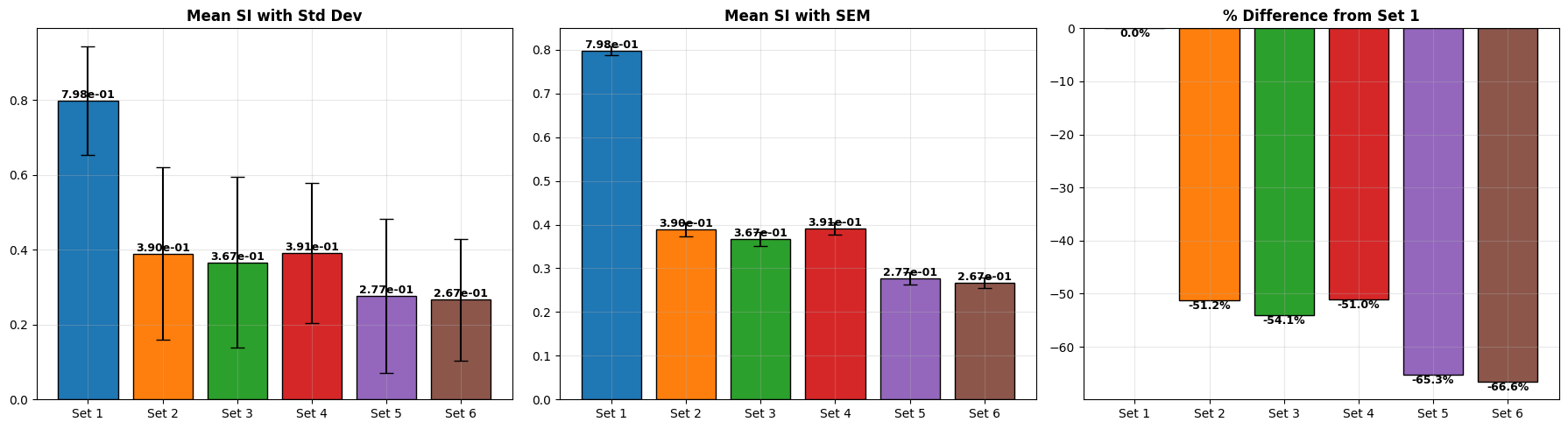}
\caption{(Left) Mean index with standard deviation; (centre) mean with standard error of the mean; (right) percentage reduction relative to Set~1.}
\label{fig:SI_stats_bar}
\end{figure}

\begin{figure}[t!]
\centering
\includegraphics[width=0.8\textwidth]{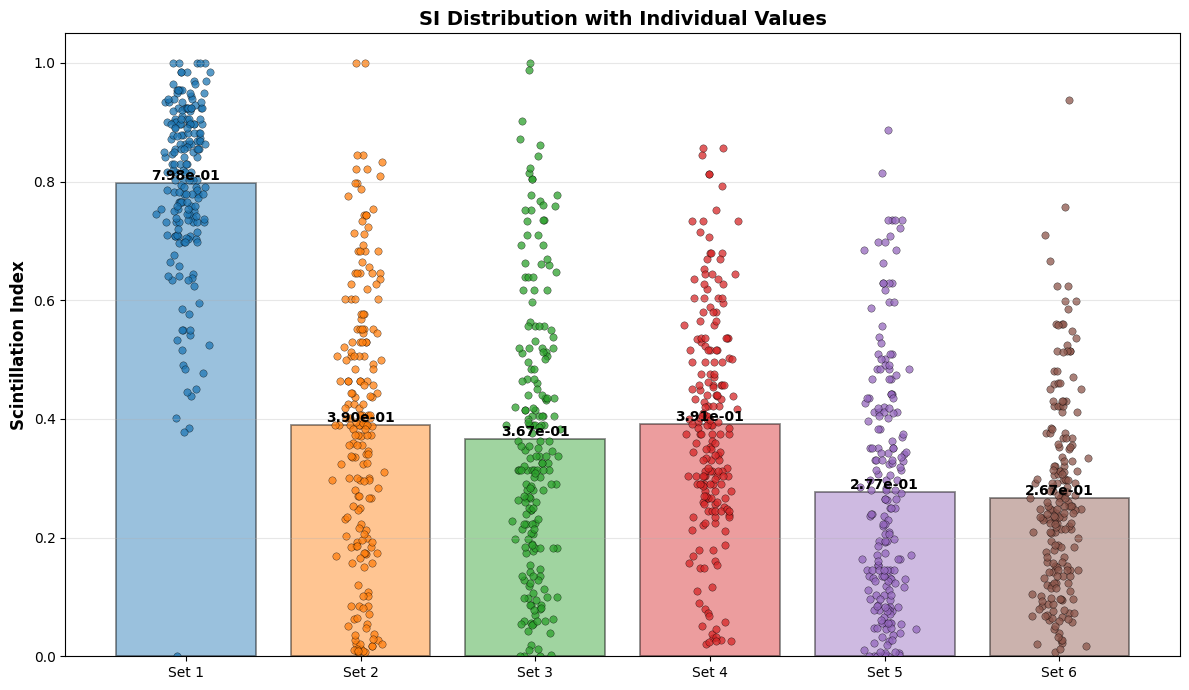}
\caption{Strip chart of per-frame pixel-level indices for Sets~1--6 with means annotated.}
\label{fig:SI_dist}
\end{figure}

The windowed index of Eq.~\eqref{eq:window} (Fig.~\ref{fig:SI_scatter}) fluctuates strongly in every set, reaching values near unity in Set~1. The set means $\pm$ standard deviation (Fig.~\ref{fig:SI_bar}) are $0.798\pm0.21$, $0.390\pm0.22$, $0.367\pm0.23$, $0.391\pm0.22$, $0.277\pm0.18$ and $0.267\pm0.15$ for $N=0$--5, corresponding to reductions of 51.2\%, 54.1\%, 51.0\%, 65.3\% and 66.6\% for $N=1$--5. The plateau at $N=1$--3, followed by a further decrease at $N=4,5$, is visible in both the SEM and the per-frame distributions (Figs.~\ref{fig:SI_stats_bar}--\ref{fig:SI_dist}). The reductions for Sets~5 and 6 relative to Set~1 exceed several SEM, but the sliding windows overlap and successive frames are correlated, so the SEM computed with 200 samples is a lower bound (Sec.~\ref{sec:limits}).

% NOTE: in the original build the two image files below carried swapped content relative to their captions;
% the filenames are assigned here so that caption and image agree (three-panel figure = Scintillation_Statistics.png,
% strip chart = Mean_Scintillation_Bar_Statistics.png). Please verify when compiling.

\begin{table}[t!]
\centering
\caption{Summary of scintillation metrics. Envelope: mean Gaussian-envelope index. Pixel: windowed index, mean $\pm$ SD. Record-level index $(\sigma/\mu)^2$ is derived from the moments annotated in Fig.~\ref{fig:histogram}.}
\label{tab:summary}
\begin{tabular}{@{}cccccc@{}}
\toprule
Set & $N$ & Envelope & Pixel (windowed) & Reduction (\%) & Record-level\\
\midrule
1 & 0 & 0.083 & $0.798\pm0.21$ & --- & 2.35\\
2 & 1 & 0.017 & $0.390\pm0.22$ & 51.2 & 0.50\\
3 & 2 & 0.049 & $0.367\pm0.23$ & 54.1 & 0.45\\
4 & 3 & 0.006 & $0.391\pm0.22$ & 51.0 & 0.34\\
5 & 4 & 0.020 & $0.277\pm0.18$ & 65.3 & 0.50\\
6 & 5 & 0.002 & $0.267\pm0.15$ & 66.6 & 0.26\\
7 & ref. & --- & --- & --- & 0.004\\
\bottomrule
\end{tabular}
\end{table}

\subsection{Comparison with the Model}

All three indicators (envelope, windowed pixel, record-level) show a clear reduction between unfiltered and five-polarizer operation, with the strongest suppression at $N=5$. Relating this to Sec.~\ref{sec:theory} requires three observations.

First, the purity relation, Eq.~\eqref{eq:purity}, concerns the total field of fully developed speckle and predicts no benefit from raising $P$. The windowed indices of Set~1 (0.80) and of the filtered sets lie below unity, so the recorded light is not fully developed Gaussian speckle; the criterion of Eq.~\eqref{eq:criterion} for a coherent core plus diffuse light is therefore the relevant description. It states that the analyzer lowers the index when the diffuse, strongly fluctuating light is predominantly cross-polarized.

Second, the geometric transversality coupling is far too weak to supply that light (cross-polarized intensity fraction $\sim\varepsilon^{4}\sim10^{-13}$, Sec.~\ref{sec:theory}). The cross-polarized diffuse component must have a material or scattering origin, such as stress birefringence or scattering in the polymer and acrylic layers of the plate. The sequential dependence on $N$ then reflects the leakage scaling $\mathrm{ER}^{-N}$ and imperfect alignment; with high-extinction elements one polarizer would saturate the effect, so the further decrease at $N=4,5$ must be confirmed with calibrated extinction ratios.

Third, the envelope also changes with $N$ (Sec.~4.2: wider fitted profile at equal wander), and a broader beam makes a fixed pixel less sensitive to the same centroid excursions. Part of the fixed-pixel reduction may therefore be geometric rather than polarization-related.

Polarimetric measurements of $\mathsf J$ before and after the plate, a crossed-analyzer control, and a wander-insensitive metric such as power in a fixed bucket would separate these contributions.

\subsection{Scope and Limitations}\label{sec:limits}

(i)~Adjacent frames are only $1.2^{\circ}$ of plate rotation apart, and Fig.~\ref{fig:gauss_stats} shows smooth, correlated drifts; the effective number of independent samples is much smaller than 200, so SEM values and significance statements should be treated as optimistic, and block-bootstrap or effective-sample-size estimates are advisable. (ii)~The tracked pixel is selected from frame~0 and that frame is retained, which biases Set~1 in particular; selecting the pixel from the time-averaged image is preferable. (iii)~The 10-frame window biases variance estimates low by a factor $(M-1)/M$ if the population variance is used, and the mean of the tracked pixel is a small fraction of full scale in several sets, so dark-offset subtraction and 8-bit quantization should be verified. (iv)~Acquisition settings and polarizer extinction ratios must be documented for each set. (v)~Matching $r_0$ does not reproduce the full strong-turbulence scintillation regime of an atmospheric path.

Within these limits, the measurements show that passive polarization filtering lowers the observed scintillation of a strongly distorted beam without sensors or feedback, and the model supplies a falsifiable criterion for when it should.

% =====================================================================
\section{Conclusion}\label{sec:conclusion}

We studied how sequential polarization filtering modifies the scintillation of a He--Ne beam transmitted through a rotating Kolmogorov-type phase plate ($r_0\approx0.6$~mm, $2w_0/r_0\approx8$) followed by zero to five co-aligned thin-film polarizers. A vector model combining the local-wave-vector transversality condition, Fresnel propagation and fourth-order moments was built from the experimental configuration. It yields $\sigma_I^{2}=(1+P^{2})/2$ for fully developed speckle, shows that the geometric polarization coupling is of order $\varepsilon^{4}\sim10^{-13}$ and thus negligible, and gives a closed criterion, Eq.~\eqref{eq:criterion}, for when filtering reduces scintillation of partially developed speckle.

Experimentally, the Gaussian-envelope index decreased from 0.083 to 0.002 (about 40-fold) and the windowed single-pixel index from $0.798\pm0.21$ to $0.267\pm0.15$ (66.6\%) between Sets~1 and 6, while beam wander stayed comparable. The decrease is not strictly monotonic in $N$, and the fixed-pixel metric is partly affected by the changing beam envelope and by correlated sampling. We attribute the reduction tentatively to rejection of a depolarized, strongly fluctuating component originating in the plate, a hypothesis that Gaussian statistics alone cannot supply and that the present data neither confirm nor exclude.

Priorities for future work are Stokes polarimetry of the plate output, calibrated extinction ratios, wander-insensitive metrics, independent-sample statistics, and extension to partially coherent and structured beams with joint optimization of coherence and polarization. The approach is passive, broadband and low-cost, which makes it a candidate element for FSO terminals operating under strong turbulence.

\section*{Funding}
Department of Science and Technology, Ministry of Science and Technology, India (CRG/2020/003338).

\section*{Declaration of Competing Interest}
Shouvik Sadhukhan reports a relationship with the Indian Institute of Space Science and Technology that includes employment. The remaining authors declare no known competing financial interests or personal relationships that could have influenced this work.

\section*{Data Availability}
The data underlying the results are available from the corresponding author upon reasonable request.

\section*{Acknowledgments}
The authors acknowledge SERB/DST (Government of India) for financial support through grant CRG/2020/003338. S.S.\ thanks the Indian Institute of Space Science and Technology, Thiruvananthapuram, for access to its laboratories.

\section*{CRediT Authorship Contribution Statement}
\textbf{Shouvik Sadhukhan:} Writing -- original draft, Visualization, Formal analysis. %\textbf{Shirsendu Sarkar:} Writing -- original draft, Visualization, Formal analysis. \textbf{Debabrata Bhadra:} Writing -- review \& editing, Validation, Supervision, Resources, Project administration, Investigation, Funding acquisition, Conceptualization.

% =====================================================================
\appendix

\section{Projection onto the Local Transverse Plane}\label{app:projection}

With $\bs=(-\varepsilon_x,-\varepsilon_y,1)/\sqrt{1+\varepsilon^{2}}$ one has $\bs\cdot\hat{\mathbf x}=-\varepsilon_x/\sqrt{1+\varepsilon^{2}}$, and
\begin{equation}
\mathbf A=\left(\bI-\bs\bs^{\mathsf T}\right)\hat{\mathbf x}
=\hat{\mathbf x}+\frac{\varepsilon_x}{1+\varepsilon^{2}}\left(-\varepsilon_x,-\varepsilon_y,1\right),
\end{equation}
which gives Eq.~\eqref{eq:Acomp}. (The sign of the $\hat{\mathbf x}$ and $\hat{\mathbf y}$ corrections is negative, as required by $\mathbf A\cdot\bs=0$.) Because $\bI-\bs\bs^{\mathsf T}$ is a projector,
\begin{equation}
|\mathbf A|^{2}=\hat{\mathbf x}^{\mathsf T}\left(\bI-\bs\bs^{\mathsf T}\right)\hat{\mathbf x}=1-(\bs\cdot\hat{\mathbf x})^{2}=1-\frac{\varepsilon_x^{2}}{1+\varepsilon^{2}}=A_x\le1.
\end{equation}
Direct expansion confirms $A_x^{2}+A_y^{2}+A_z^{2}=(1+\varepsilon_y^{2})/(1+\varepsilon^{2})=A_x$. The projection thus removes, rather than redistributes, the fraction $1-A_x$ of the trial intensity. In the paraxial limit $\varepsilon\ll1$,
\begin{equation}
\mathbf A\simeq\left(1-\varepsilon_x^{2},\;-\varepsilon_x\varepsilon_y,\;\varepsilon_x\right),\qquad
\bE_t\simeq\mathcal E\left[\hat{\mathbf x}+\varepsilon_x\hat{\mathbf z}\right]\ \text{to first order},
\end{equation}
which establishes the hierarchy: scalar phase modulation at zeroth order, the longitudinal field at first order in $\varepsilon$, and cross-polarization coupling at second order.

\section{Component-wise Fresnel Integrals and Second-Order Correlation}\label{app:fresnel}

Writing $R^{2}=|\bR-\brho|^{2}$, the three components of Eq.~\eqref{eq:fresnel} are
\begin{align}
E_x(\bR,z)&=\frac{E_0e^{i\kappa z}}{i\lambda z}\iint e^{-|\brho|^{2}/w_0^{2}}e^{i\Phi(\brho)}\left(1-\frac{\varepsilon_x^{2}}{1+\varepsilon^{2}}\right)e^{\frac{i\kappa}{2z}R^{2}}\,d^{2}\rho,\\
E_y(\bR,z)&=-\frac{E_0e^{i\kappa z}}{i\lambda z}\iint e^{-|\brho|^{2}/w_0^{2}}e^{i\Phi(\brho)}\frac{\varepsilon_x\varepsilon_y}{1+\varepsilon^{2}}\,e^{\frac{i\kappa}{2z}R^{2}}\,d^{2}\rho,\\
E_z(\bR,z)&=\frac{E_0e^{i\kappa z}}{i\lambda z}\iint e^{-|\brho|^{2}/w_0^{2}}e^{i\Phi(\brho)}\frac{\varepsilon_x}{1+\varepsilon^{2}}\,e^{\frac{i\kappa}{2z}R^{2}}\,d^{2}\rho.
\end{align}
The scalar counterpart omits $A_j$. The mutual coherence at the same detector point is
\begin{equation}
\Gamma^{(1,1)}_{x}(\bR)=\frac{E_0^{2}}{(\lambda z)^{2}}\iiiint e^{-(|\brho_1|^{2}+|\brho_2|^{2})/w_0^{2}}
\avg{A_x(\brho_1)A_x(\brho_2)\,e^{i[\Phi(\brho_1)-\Phi(\brho_2)]}}
e^{\frac{i\kappa}{2z}\left(R_1^{2}-R_2^{2}\right)}d^{2}\rho_1d^{2}\rho_2 ,
\end{equation}
where the joint average must be retained because $A_x$ and $\Phi$ both depend on $\phi_t$. Since $A_x=1-O(\varepsilon^{2})$, to leading order the average is $\avg{e^{i[\Phi(\brho_1)-\Phi(\brho_2)]}}=\exp\!\left[-\tfrac12D_\phi(|\brho_1-\brho_2|)\right]=\exp\!\left[-3.44\,(|\brho_1-\brho_2|/r_0)^{5/3}\right]$ for a Gaussian phase with the Kolmogorov structure function. The mean intensity is $\avg{I_x}=\Gamma^{(1,1)}_x(\bR)$.

\section{BCP Matrix, Fourth-Order Moments and the Purity Relation}\label{app:purity}

The beam coherence--polarization (BCP) matrix is $\Gamma^{(1,1)}_{\alpha\beta}(\brho_1,\brho_2,z)=\avg{E_\alpha^{*}(\brho_1,z)E_\beta(\brho_2,z)}$, $\alpha,\beta\in\{x,y\}$, which reduces to the polarization matrix $\mathsf J(\brho)$ for $\brho_1=\brho_2$ \cite{korotkova2006,korotkova2008}. In a linear isotropic medium the orthogonal components propagate independently, so each element obeys the scalar propagation law.

For a zero-mean circular complex Gaussian vector, the moment theorem \cite{reed} gives $\avg{E_i^{*}E_j^{*}E_kE_l}=\mathsf J_{ik}\mathsf J_{jl}+\mathsf J_{il}\mathsf J_{jk}$. Hence $\Gamma^{(2,2)}_{iijj}=\avg{I_iI_j}=\mathsf J_{ii}\mathsf J_{jj}+|\mathsf J_{ij}|^{2}$, so that $\sigma_{ii}^{2}=1$ and $\sigma_{xy}^{2}=|\mathsf J_{xy}|^{2}/(\mathsf J_{xx}\mathsf J_{yy})$. Inserting these in Eq.~\eqref{eq:SI},
\begin{equation}
\sigma_I^{2}=\frac{\mathsf J_{xx}^{2}+\mathsf J_{yy}^{2}+2|\mathsf J_{xy}|^{2}}{(\mathsf J_{xx}+\mathsf J_{yy})^{2}}=\frac{\tr(\mathsf J^{2})}{[\tr\mathsf J]^{2}}.
\end{equation}
For a $2\times2$ Hermitian matrix $\tr(\mathsf J^{2})=[\tr\mathsf J]^{2}-2\det\mathsf J$, and the degree of polarization obeys $P^{2}=1-4\det\mathsf J/[\tr\mathsf J]^{2}$. Therefore
\begin{equation}
\sigma_I^{2}=1-\frac{2\det\mathsf J}{[\tr\mathsf J]^{2}}=1-\frac{1-P^{2}}{2}=\frac{1+P^{2}}{2},
\end{equation}
which is Eq.~\eqref{eq:purity}. The values $1/2$ (unpolarized) and $1$ (fully polarized) reproduce the factor of two reported in Ref.~\cite{korotkova2006}.

\section{Two-Component Model of Partially Developed Speckle}\label{app:twocomp}

Let $E_x=s+g_x$, $E_y=g_y$ with $s$ deterministic ($S=|s|^{2}$) and $g_x,g_y$ independent circular Gaussian variables with powers $d_x,d_y$. For $I_x=|s+g_x|^{2}$ (Rician statistics), $\avg{I_x}=S+d_x$ and $\mathrm{Var}(I_x)=d_x^{2}+2Sd_x$, while $I_y=|g_y|^{2}$ has $\avg{I_y}=d_y$, $\mathrm{Var}(I_y)=d_y^{2}$. By independence the total intensity has mean $S+d_x+d_y$ and variance $d_x^{2}+d_y^{2}+2Sd_x$, which gives Eq.~\eqref{eq:twocomp}. The inequality $\sigma_{\rm tot}^{2}>\sigma_x^{2}$ with $u=S+d_x$ and $a=d_x(d_x+2S)$ reads $(a+d_y^{2})u^{2}>a(u+d_y)^{2}$, i.e.\ $d_y(u^{2}-a)>2au$. Since $u^{2}-a=S^{2}$, this is $d_yS^{2}>2d_x(d_x+2S)(S+d_x)$, i.e.\ Eq.~\eqref{eq:criterion}. For $S\to0$ the threshold diverges and polarization filtering cannot reduce the index, in agreement with the Gaussian result $\sigma_{xx}^{2}=1$. For $N$ imperfect analyzers, replacing $d_y$ by $\eta^{2N}d_y$ gives
\begin{equation}
\sigma_N^{2}=\frac{d_x^{2}+\eta^{4N}d_y^{2}+2Sd_x}{(S+d_x+\eta^{2N}d_y)^{2}}.
\end{equation}

% =====================================================================

\end{document}